\documentclass[a4paper,11pt]{article}
\usepackage{pos}

\title{Continuum limit of generalized charm susceptibilities}
\author*[a]{Sipaz Sharma}

\affiliation[a]{
  Physik Department, Technische Universit\"at M\"unchen,\\ James-Franck-Stra{\ss}e 1, D-85748 Garching~b.~M\"unchen, Germany}
\note{For the HotQCD Collaboration.}

\emailAdd{sipaz.sharma@tum.de}
\abstract{We investigate in detail the sensitivity of the fourth-order charm fluctuation, calculated on the lattice, to the input bare charm quark mass. We approach its continuum limit by employing four different lines of constant physics. We quantify the cutoff effects arising due to bare charm quark mass for both coarser and finer lattices. Finally, we show that the ratios of generalized charm susceptibilities, calculated on coarser lattices, and the continuum-extrapolated fourth-order charm fluctuation are sufficient to obtain the continuum limit of all physical observables relevant to open charm physics. }

\FullConference{The 41st International Symposium on Lattice Field Theory (LATTICE2024)\\
 28 July - 3 August 2024\\
Liverpool, UK\\}

\begin{document}
\maketitle

\section{Introduction}
Generalized charm susceptibilities are a powerful tool for probing the nature of charm degrees of freedom at finite temperature, $T$,  using lattice QCD techniques. These are derivatives of the QCD pressure, $P$, with respect to chemical potentials corresponding to the conserved charges: baryon number ($B$), electric charge ($Q$), strangeness ($S$)
and charm ($C$), \begin{equation} 
	{\chi^{BQSC}_{klmn}=\dfrac{\partial^{(k+l+m+n)}\;\;[P\;(\hat{\mu}_B,\hat{\mu}_Q,\hat{\mu}_S,\hat{\mu}_C)\;/T^4]}{\partial\hat{\mu}^{k}_B\;\;\partial\hat{\mu}^{l}_Q\;\;\partial\hat{\mu}^{m}_S\;\;\partial\hat{\mu}^{n}_C}}\bigg|_{\vec{\mu}=0} .\; 
	\label{eq:chi}
\end{equation}
Here, we introduce a dimensionless notation for chemical potentials, 
${\hat{\mu}_X = \mu_X/T}$, with $X \in \{B, Q, S, C\}$.
  Our previous works \cite{Sharma:2022ztl,BAZAVOV2024138520,Sharma:2024ucs, Sharma:2024edf,Sharma:2025zhe} have established that below chiral crossover temperature, ${T_{pc}=156.5\pm1.5}$ MeV \cite{HotQCD:2018pds}, hadrons are the carriers of the charm quantum number, $C$, whereas at $T_{pc}$ charm deconfinement sets in leading to the emergence of charm quarks as new degrees of freedom. However, the disappearance of charmed hadrons is gradual, and charm quarks give dominant contribution to the partial charm pressure  only above $175$ MeV. Additionally, lattice investigations of charm susceptibilities have also predicted the existence of experimentally unobserved charmed hadrons in the low temperature phase \cite{Bazavov:2014yba,BAZAVOV2024138520}. As pointed out in one of our previous articles, \cite{Sharma:2025zhe}, above conclusions are based on the analysis of the ratios of various linear combinations of the generalized charm susceptibilities. These linear combinations are determined by the relevant physics which one aims to probe. These ratios were calculated on coarser lattices (temporal lattice extent, $N_\tau=8$) but the conclusions driven from their analysis hold in the continuum limit because the dominant cutoff effects present due to a relatively heavier charm quark mass cancel in the ratios. However, ratios alone are not sufficient to fully understand charm thermodynamics. Therefore, taking the continuum limit of the proxies for the partial pressures of the relevant charm degrees of freedom is crucial. These proxies are constructed from linear combinations of various generalized charm susceptibilities. In the next sections, a summary of the approach to continuum limit of the least noisy charm susceptibility ($\chi^{BQSC}_{0004} \equiv \chi^{C}_{4}$) is provided -- full details will be provided in our forthcoming publication. We argue that only continuum extrapolated $\chi^{C}_{4}$  is enough to convert observables, which are normalized by $\chi^{C}_{4}$ and computed on coarser lattices,  to their absolute continuum values.
  \section{Lattice setup}
  We used (2+1)-flavor HotQCD configurations generated using HISQ action and a Symanzik-improved gauge action for physical strange-to-light quark mass ratio, ${m_s/m_{l}}=27$, for three temporal lattice extents, ${N_{\tau}}=8, 12 \text{ and } 16$. The calculation of derivatives of the QCD pressure was achieved by the unbiased stochastic estimation of various traces using random noise method. In particular, 500 random vectors were used to calculate traces relevant for  $\chi^{C}_{4}$ on each configuration. The temperature scale was determined using a parametrization of the kaon decay constant in lattice units, $af_K$, as given in \cite{Bollweg:2021vqf}, and the conversion to physical units was performed using the $f_K$ value given in the latest FLAG review \cite{FlavourLatticeAveragingGroupFLAG:2024oxs}. Since the temperature is given by $T=(aN_\tau)^{-1}$, this implies that we used three different lattice spacings at a fixed temperature. 
   \begin{figure}[]	
  	\includegraphics[width=\linewidth]{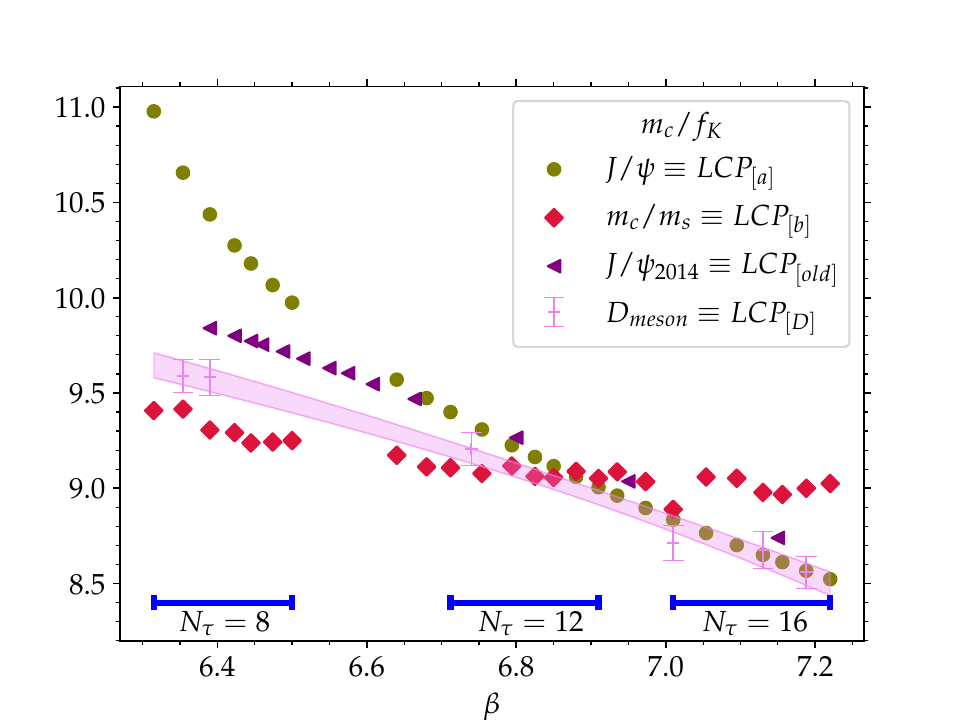}
  	\caption{Shown are the bare charm quark mass values normalised by $af_{K}$ as a function of the inverse gauge coupling, $\beta$, tuned using four different criteria to define lines of constant physics: LCP$_{[a]}$, LCP$_{[b]}$, LCP$_{[old]}$, and LCP$_{[D]}$ (see text for definitions). The band shows bootstrap error of LCP$_{[D]}$ values. The blue lines explicitly show the ranges of inverse gauge couplings, $\beta$, relevant for three different temporal lattice extents used in this work.}
  	\label{fig:para_comp}
  \end{figure}
  
  We treated charm quark in the quenched approximation, and used the so-called epsilon term to remove $\mathcal{O}((am_c)^4)$ tree-level lattice artifacts \cite{Follana:2006rc}. In order to understand the cutoff effects in $\chi^{C}_{4}$  due to input bare charm quark mass, $am_c$, we used $am_c$ values tuned on four different Lines of Constant Physics (LCPs) shown in Fig.~\ref{fig:para_comp} as functions of the inverse gauge coupling, $\beta$:\\
  $\bullet$ {LCP$_{[a]}$} was tuned by keeping spin-averaged charmonium mass fixed to its physical value. For further details, see \cite{Sharma:2022ztl}.\\
  $\bullet$  {LCP$_{[b]}$} was tuned by keeping the ratio $am_c/am_s$ fixed to its Particle Data Group (PDG) value, $11.76$ \cite{ParticleDataGroup:2024cfk}. This LCP is based on a strange quark mass tuned by fixing the mass of the fictitious pseudoscalar meson, $\eta_{s\bar{s}}$, to $695$ MeV \cite{Bazavov:2011nk}. However, this is slightly higher than the value given by  $\chi$PT, $686$ MeV, indicating that the strange quark mass is $2.6\%$ larger than its physical value. In addition to this, for finer lattices corresponding to $\beta>7.03$, the resulting lattice mass of $\eta_{s\bar{s}}$ is larger than  $695$ MeV by about $3.5\%$. This drift from $695$ MeV was further corrected by using lowest order $\chi$PT such that $M_{\eta_{s\bar{s}}}^{2}\propto m_s$ (see Ref.~\cite{HotQCD:2014kol}) but the configuration generation did not take into account this corrected version of LCP. This mistuning is clearly reflected in the $am_c$ values corresponding to LCP$_{[b]}$ in the higher $\beta$ range of Fig.~\ref{fig:para_comp}.\\
  $\bullet$ {LCP$_{[old]}$} is an older prescription used in Ref.~\cite{Bazavov:2014yba}. It also was tuned by keeping spin-averaged charmonium mass fixed to its physical value.\\
  $\bullet$ $am_c$ values on  LCP$_{[D]}$ correspond to physical  D-meson mass.   $am_c$ parametrization on this LCP will be given in our forthcoming publication.\\
  In the following, results based on above LCPs will carry subscripts $[a]$,$[b]$,$[old]$ and $[D]$ respectively. 
  \section{Fourth-order charm fluctuation on different LCPs}
  \begin{figure}[]
  	\includegraphics[width=0.48\linewidth]{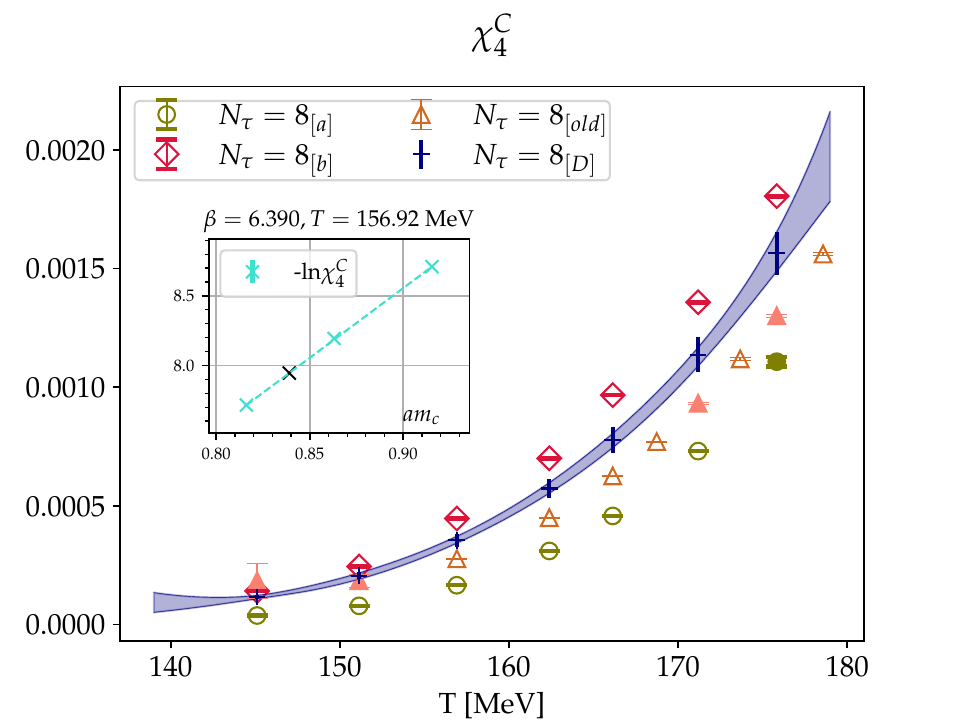}
  		\includegraphics[width=0.48\linewidth]{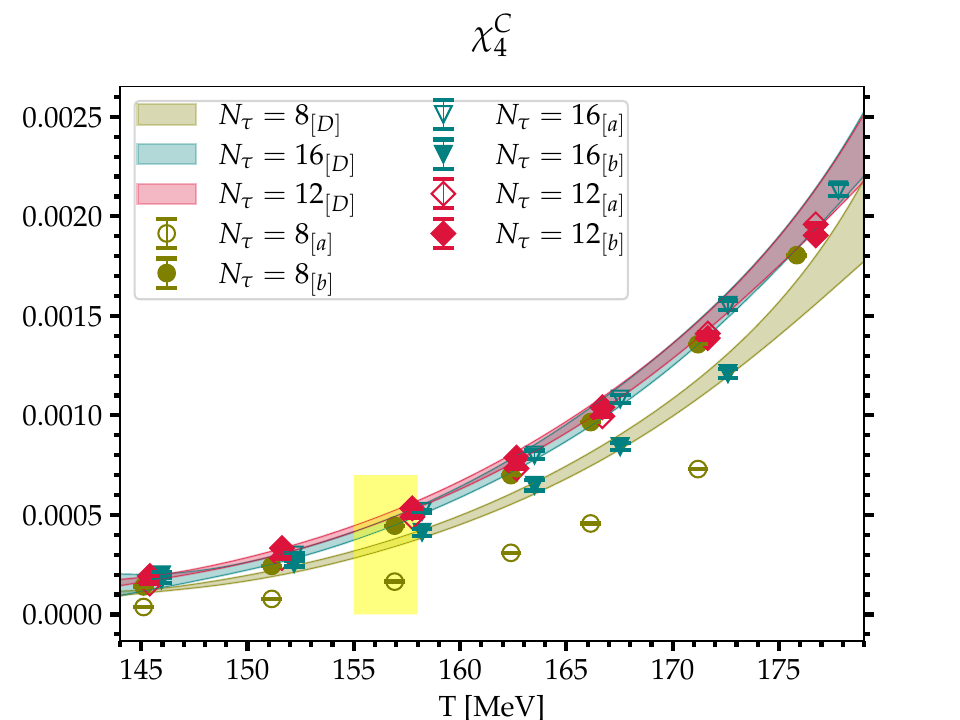}
  	\caption{Shown is $\chi^{C}_{4}$  as a function of temperature constructed on LCP$_{[D]}$ using its calculated values on LCP$_{[a]}$, LCP$_{[b]}$ and LCP$_{[old]}$    for $N_\tau=8$ lattices. The solid points and the bands shown are results of $[2, 2]$ Pad\'e interpolations. The inset  shows linear interpolation of $-\textrm{ln } \chi^C_4$ in $am_c$ at $\beta=6.390$ and the black cross represents $-\textrm{ln } \chi^C_4$ on LCP$_{[D]}${\it (Left)}. Lattice QCD results for 
  		$\chi_4^C$ for different values of the temporal lattice extents calculated on the two different LCPs, {\it i.e.} LCP$_{[a]}$ and LCP$_{[b]}$, and constructed on  LCP$_{[D]}$. $\chi_4^C$ for $N_\tau=16$ on LCP$_{[D]}$ represents our continuum estimate. The yellow band represents $T_{pc}$ with its uncertainty {\it (Right)}. }
  	\label{fig:LCP_D}
  \end{figure}
   In the hadron gas phase below $T_{pc}$, at a temperature, $T$, contribution to the partial charm pressure from each charmed state of mass, $m_i$ carrying quantum numbers, $B_i$, $Q_i$,$S_i$, $C_i$, is proportional to,
   \begin{equation}
   \bigg(\dfrac{m_i}{T}\bigg)^{3/2}e^{-m_i/T}\;[1+\mathbb{O}((m_i/T)^{-1})]\text{cosh}(B_i\hat{\mu}_B+Q_i\hat{\mu}_Q+S_i\hat{\mu}_S+C_i\hat{\mu}_C).
   \label{eq:contribution}
    \end{equation} 
This also holds true in the QGP phase where charm quarks and charmed hadrons coexist \cite{BAZAVOV2024138520}. Therefore, the exponential suppression in \eqref{eq:contribution} dictates that the lightest charmed state should give the dominant contribution to the partial charm pressure. In the hadron gas phase, the lighest charmed state is D-meson. In the QGP phase, slightly above $T_{pc}$, thermal mass of charm quark-like quasi particle is around D-meson mass \cite{Sharma:2025zhe}. Therefore, based on thermodynamical arguments, $\chi^C_4$ on LCP$_{[D]}$ will be closer to the physical case, particularly at relatively coarser  $N_{\tau}=8$ lattices. As can be inferred from Fig.~\ref{fig:para_comp}, $am_c$ at the lowest $\beta$ value, $6.315$, used in the calculation of  $\chi^C_4$ at $N_\tau=8$, varies by 16 \% between  LCP$_{[a]}$ and  LCP$_{[b]}$, whereas for the lowest  $\beta$ value of $N_\tau=12$ calculation, 6.712,  this variation reduces to $3\%$. However, as discussed in the previous section, due to mistuning of $am_s$,  LCP$_{[b]}$ drifts away from  LCP$_{[a]}$ for $\beta > 6.9$. Nonetheless,  LCP$_{[a]}$ and   LCP$_{[D]}$ converge in the higher $\beta$  range, which is relevant for the finest $N_\tau=16$ lattices.  

 The mass of  lightest charmed state is proportional to $am_c$, and since charm quark mass is an order of magnitude larger than the temperature of interest, thus  even a small change in the charm quark mass can lead to large changes in the Boltzmann weight in \eqref{eq:contribution}. This sensitivity of the partial charm pressure to $am_c$ is clearly reflected in its fourth derivative w.r.t. $\hat{\mu}_{C}$ shown in Fig.~\ref{fig:LCP_D}  [Left] for $N_\tau=8$ lattices. For  $N_\tau=8$, we used lattice calculations of  $\chi^C_4$ on three LCPs: LCP$_{[a]}$, LCP$_{[b]}$, LCP$_{[old]}$ to construct $\chi^C_4$ on LCP$_{[D]}$. As shown in the inset of Fig.~\ref{fig:LCP_D}  [Left] , at each temperature, we linearly interpolated $-\ln\chi^C_4$ in $am_c$ to obtain $\chi^C_4$ on LCP$_{[D]}$. Similarly, we constructed $\chi^C_4$ on LCP$_{[D]}$ for  $N_\tau=12$ and $16$. Futher details, especially on error analysis, will be given in a forthcoming publication. 
  \section{Continuum limit}

 \begin{figure}[]
		\includegraphics[width=0.45\linewidth]{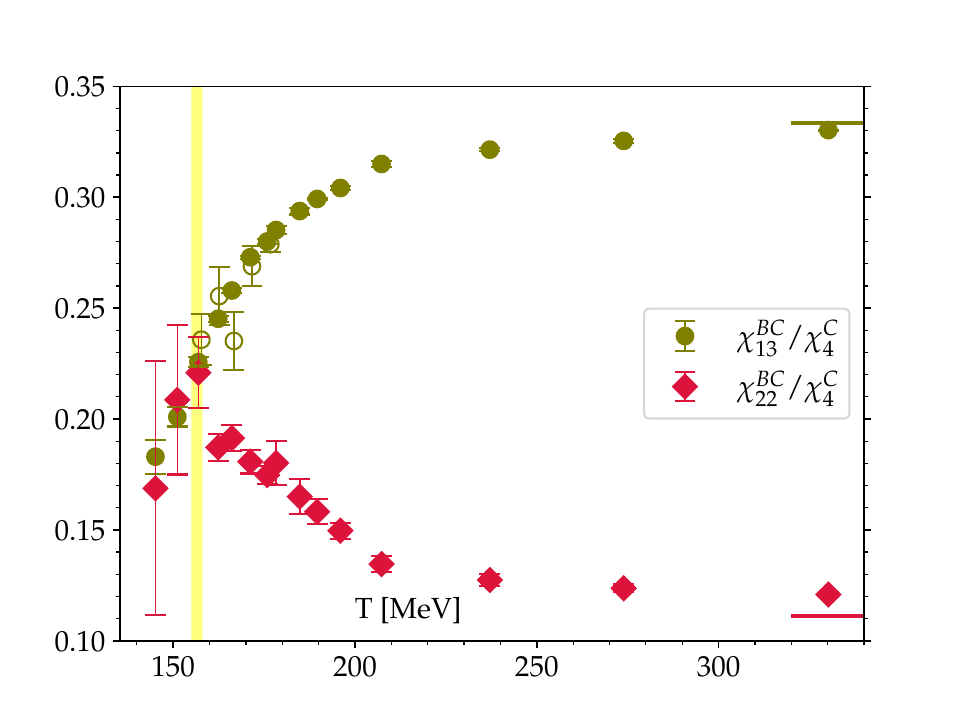}
		\includegraphics[width=0.45\linewidth]{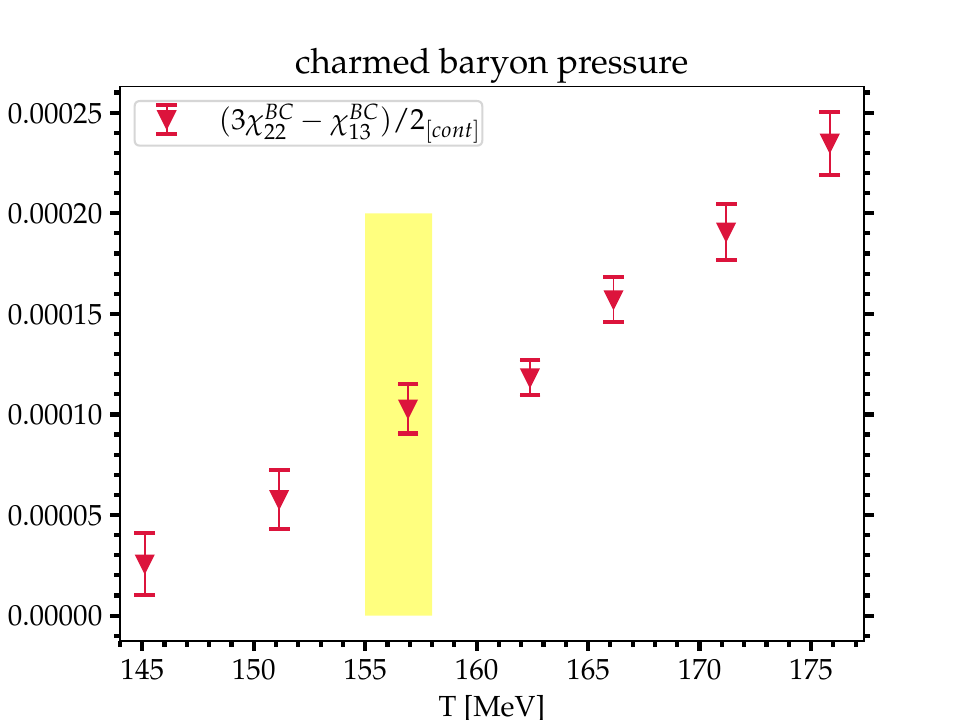}
	\caption{The ratios $\chi_{mn}^{BC}/\chi_4^C$ as function of the temperature
		compared to the ideal quark gas predictions at high temperatures shown as horizontal lines.  Solid markers represent $N_{\tau}=8_{[b]}$ results, whereas unfilled markers of the same color represent the respective $N_{\tau}=12_{[b]}$ results. Solid markers for $T>176$~MeV represent $N_{\tau}=8_{[old]}$ results taken from Ref. \cite{Bazavov:2014yba}.  The vertical yellow band represents chiral crossover temperature with its uncertainty {\it (Left)}. Shown is the continuum result of a proxy for charmed baryon partial pressure  as function of temperature. The vertical yellow band represents chiral crossover temperature with its uncertainty {\it (Right)}. }
	\label{fig:ratio}

\end{figure}

	We show lattices calculations of $\chi_4^C$ on LCP$_{[a]}$ and  LCP$_{[b]}$ for three $N_\tau$ values in Fig.~\ref{fig:LCP_D}~ [Right]. The figure also shows $\chi_4^C$ constructed on LCP$_{[D]}$ for three $N_\tau$ values. As expected, with the convergence of $am_c$ on   LCP$_{[a]}$ and LCP$_{[D]}$,  $\chi_4^C$ also converges on  LCP$_{[a]}$ and  LCP$_{[D]}$ for $N_\tau>8$.  As can be seen from Fig.~\ref{fig:LCP_D}~ [Right], the $\chi_4^C$ bands corresponding to LCP$_{[D]}$  agree within errors for $N_\tau=12$ and $16$. Therefore, we use  $N_\tau=16$ result as our continuum estimate of $\chi_4^C$.

It was shown in our previous work \cite{BAZAVOV2024138520} that at a fixed $N_\tau$, sensitivity to the choice of LCP cancels to a large extent in the  ratios of generalized susceptibilities. In Fig.~\ref{fig:ratio}~[Left], two $N_\tau=8$ baryon-charm correlations normalised by $\chi_4^C$  are shown. Additionally,  agreement of $N_\tau=8_{[b]}$ and  $N_\tau=12_{[b]}$ results of $\chi^{BC}_{13}/\chi_4^C$  in Fig.~\ref{fig:ratio}~[Left] implies that the cutoff effects cancel to a large extent in the ratios. Therefore, physical conclusions drawn from ratios calculated on the coarser lattices hold in the continuum limit. This implies that by multiplying $N_\tau=8_{[b]}$ version of observables normalised by $\chi_4^C$ to the continuum version of $\chi_4^C$ and after doing error propagation, one can obtain continuum limit of the relevant observables.  Fig.~\ref{fig:ratio}~[Right] shows a linear combination of the unnormalised  versions of baryon-charm correlations of Fig.~\ref{fig:ratio}~[Left]. According to the quasi-particle model from Ref.~\cite{BAZAVOV2024138520}, $(3\chi^{BC}_{22}-\chi^{BC}_{13})/2$ is a proxy for the charmed baryon pressure, and Fig.~\ref{fig:ratio}~[Right] shows its continuum result. Similarly, continuum limit of other physical quantities can also be obtained by following above procedure.
  \section{Conclusions and Summary}
  We showed that the generalized charm susceptibilities are sensitive to the choice of LCPs used to tune the bare charm quark mass.  In order to reduce cutoff effects for coarser lattices at finite temperature, the choice of LCP should be motivated by the thermodynamics. On the other hand, for finer lattices, correctly tuned LCPs converge.  By quantifying  the cutoff effects arising due to bare charm quark mass in the fourth-order charm fluctuation, we obtained its continuum limit. Finally, using the continuum extrapolated  $\chi_4^C$, we indirectly obtained continuum results of other physical quantities essential for describing charm thermodynamics. As an example, we showed continuum extrapolated charmed baryon pressure.
\section*{Acknowledgments}

This work was supported by The Deutsche Forschungsgemeinschaft (DFG, German Research Foundation) - Project number 315477589-TRR 211,
``Strong interaction matter under extreme conditions”. 
The authors gratefully acknowledge the
computing time and support provided to them on the high-performance computer Noctua 2 at the NHR Center
PC2 under the project name: hpc-prf-cfpd. These are funded by the Federal Ministry of Education
and Research and the state governments participating on the basis of the resolutions of the GWK
for the national high-performance computing at universities (www.nhr-verein.de/unsere-partner).
Numerical calculations have also been performed on the
GPU-cluster at Bielefeld University, Germany. We thank the Bielefeld HPC.NRW team for their support. 

All computations in this work were performed using \texttt{SIMULATeQCD} code ~\cite{HotQCD:2023ghu}. All the HRG calculations were performed using the AnalysisToolbox code developed by the HotQCD Collaboration \cite{Clarke:2023sfy}.

\bibliographystyle{JHEP}
\bibliography{refs}

\end{document}